\documentclass[12pt]{elsarticle}
\usepackage{graphicx}
\usepackage{amssymb}

\def\gr{$\gamma$-ray}

\begin{document}
\begin{frontmatter}
\title{Remote sensing of clouds and aerosols with cosmic rays} 

\author{Andrii Neronov}
\ead{Andrii.Neronov@unige.ch, Tel: +41223792120}
\address{ISDC Data Centre for Astrophysics, Astronomy Department, Geneva University
Ch. d'Ecogia 16, Versoix,1290, Switzerland}
\author{Denys Malyshev}
\address{ISDC Data Centre for Astrophysics, Astronomy Department, Geneva University
Ch. d'Ecogia 16, Versoix,1290, Switzerland}
\author{Anton Dmytriiev}
\address{Physics Department, Taras Shevchenko National University of Kyiv, 
Glushkova Av. 2, 03022, Kiev, Ukraine}

\begin{abstract}
Remote sensing of atmosphere is conventionally done via a study of extinction / scattering of light from natural (Sun, Moon) or artificial (laser) sources. Cherenkov emission from extensive air showers generated by cosmic rays provides one more natural light source distributed throughout the atmosphere. We show that Cherenkov light carries information on three-dimensional distribution of clouds and aerosols in the atmosphere and on the size distribution and scattering phase function of cloud/aerosol particles. Therefore, it could be used for the atmospheric sounding. The new atmospheric sounding method  could be implemented via an adjustment of technique of imaging Cherenkov telescopes. The atmospheric sounding data collected in this way could be used both for atmospheric science and for the improvement of the quality of astronomical gamma-ray observations. 
 \end{abstract}

\end{frontmatter}

\section{Introduction}

Knowledge of optical properties of clouds and aerosols is important in a wide range of scientific problems, from atmospheric and climate science \cite{ipcc13} to astronomical observations across wavelength bands \cite{beniston02,font12,chaves13}. 

Clouds are reflecting and absorbing radiation form the Sun, thus regulating the intake of the Solar energy by the Earth. Study of scattering and absorption of light by clouds is, therefore, a key element for understanding of the physics of the Earth atmosphere \cite{ipcc13,stephens05}. Aerosols work as condensation centres for formation of cloud water droplets and ice crystals. Understanding of relation between clouds and aerosols is one of the main challenges of atmospheric science \cite{ipcc13,haywood00}.   

Probes of the properties of clouds and aerosols are done using in situ measurements and remote sensing techniques \cite{stephens07} including imaging  from space or from the ground \cite{king92}, observations of transmitted light from the Sun or Moon  \cite{bovensmann99} and sounding of the clouds with radiation beams \cite{winker09}. LIght Detection And Ranging (LIDAR) sounding techniques (Fig. \ref{fig:principle}) probe vertical structure of clouds and aerosols via timing of backscatter signal from a laser beam  \cite{winker09}. 

Presence of clouds perturbs astronomical observations in the Very-High-Energy (VHE) \gr\ (photons with energies 0.1-10~TeV) band and operation of Cosmic Ray (CR) experiments which use the Earth atmosphere as a giant high-energy particle detector \cite{font12,chaves13}. Imaging Atmospheric Cherenkov Telescope (IACT) arrays\footnote{HESS telescopes: http://www.mpi-hd.mpg.de/hfm/HESS/; MAGIC telescopes: https://magic.mpp.mpg.de; VERITAS telescopes: http://veritas.sao.arizona.edu.}, as well as air fluorescence telescopes for detection of Ultra-High-Energy CRs\footnote{Pierre Auger Observatory: http://www.auger.org;  Telescope Array: http://www.telescopearray.org; JEM-EUSO: http://jemeuso.riken.jp/en/.} detect cosmic high-energy particles via imaging of Cherenkov and fluorescence emission from the particle Extensive Air Showers (EAS), initiated by the primary cosmic particles. Information on the presence and properties of the clouds and aerosols is essential for the proper interpretation of the data collected in this way. Gamma-ray  / CR observations affected even by optically thin clouds  are normally excluded from data sets, because the properties of the clouds are not known sufficiently well to allow correction for the effects of scattering  of light by the atmospheric features. 

Here we show that Cherenkov light produced by the  EAS could be used as a tool for remote sensing of the atmosphere. We show that this tool allows characterisation of three-dimensional cloud / aerosol coverage above the observation site and provides information on  physical properties of  cloud and aerosol particles. 

\section{The remote sensing method}

CRs of energy $E>E_{cr}$ are hitting the atmosphere from all directions at a rate \cite{pdd}
$N_{cr}(E_{cr})\sim 10\left[E_{cr}/100\mbox{ GeV}\right]^{-\gamma_{cr}+1}(\mbox{ m}^2\mbox{ s sr})^{-1}$ where $\gamma_{cr}\simeq 2.7$ is the slope of the CR spectrum.
Each CR induced EAS generates a short flash of visible-to-UV light via Cherenkov and fluorescence mechanisms. The entire set of EAS provides a source of light distributed everywhere throughout the atmospheric volume.  This light source is well calibrated since the CR flux is well measured and is almost invariable in time at the energies above 100~GeV \cite{pdd}. Techniques for detection of light flashes from EAS are nowadays well developed and are widely used in \gr\ astronomy \cite{aharonian_book} and CR research \cite{grieder_book}. It is, therefore, natural to consider a possibility to use the light from EAS for the study of the atmosphere. 

\begin{figure}
\includegraphics[width=0.6\linewidth]{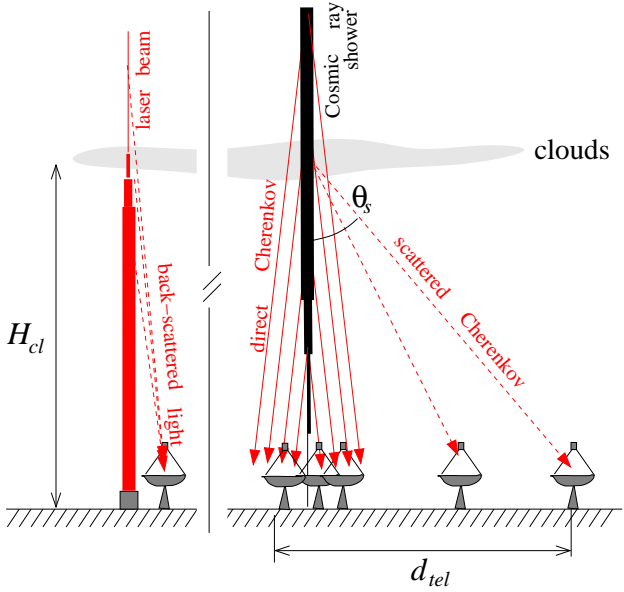}
\caption{Principle of remote sensing of the atmosphere with Cherenkov light beams of EAS (right) shown in comparison with the scheme of operation of LIDARs (left).  }
\label{fig:principle}
\end{figure}

The suggested principle of remote sensing of the atmosphere  with EAS light is shown in Fig. \ref{fig:principle}.  The EAS  is observed from the ground by a system of (minimum two) telescopes. Stereo vision technique enables reconstruction of three-dimensional geometry of the EAS and measurement of the number of photons from each EAS as function of altitude, $dN_\gamma/dH$. Measurements for individual EAS are combined into a cumulative vertical profile $dN_{\gamma,cum}/dH(x,y,H)$ of emission from a large number of EAS. It is a function of the three-dimensional position in the atmospheric volume, $(x,y,H)$ within the Field-of-View (FoV) of the telescopes. Scattering and absorption of light by atmospheric features leads to the distortion of the cumulative vertical profile. Characterisation of these distortions provides a tool for the measurement of the distribution and optical properties of the features. 

\section{Vertical profile of Cherenkov light in clear and cloudy sky}

High-energy particles forming an EAS move with velocities faster than the speed of light in the atmosphere. Superluminal motion is accompanied by emission of Cherenkov radiation at a known rate $Y_{Ch}$ \cite{hillas82}. The number of high-energy particles in the EAS is, on average, also a known function $N_p(E_{cr},X)$ of the atmospheric column density along the EAS  track,  $X$ \cite{hillas}. The overall flux of Cherenkov light from the EAS is the sum of emission from all the particles,
$dN_{Ch}/dX(E_{cr},X,\alpha,\lambda)=Y_{Ch}N_p(E_{cr},X)$. It depends on the photon wavelength, $\lambda$ and  on the angle between the EAS axis and photon direction, $\alpha$.  Cherenkov photons are concentrated in a narrow beam with approximately exponential profile $dN_{Ch}/dX\sim \exp\left(-\alpha/\alpha_c\right)$, where  $\alpha_c\sim 1^\circ$ is a cut-off angle determined by the angle of Cherenkov emission in the air and by the angular spread of high-energy particles \cite{nerling06}. Linear brightness profile of EAS as a function of altitude $H$ is $dN_{Ch}/dH =(\rho/\cos\theta_z) dN_{Ch}/dX$ where $\rho$ is the atmospheric density and $\theta_z$ is the zenith angle of the EAS. In our illustrative example we consider observations in Zenith direction, $\theta_z\simeq 0^\circ$. 

Attenuation of visible and UV light during propagation through the atmosphere  determines the number of photons $N_\gamma$ reaching the telescope 
 \begin{equation}
 \label{eq:shower_eq}
 \frac{dN_{\gamma}}{dH}=\frac{dN_{Ch}(H,\alpha(H),\lambda)}{dH}\exp\left(-\tau(H,\lambda)\right)
 \end{equation}
 where the $\alpha(H)$ is the angle at which the EAS track is visible from the location of the telescope. The optical depth $\tau$ is determined by the density $n_s$ and extinction cross-section $\sigma_s$ of  the scattering centres, $\tau=\int_{H_{tel}}^{H} \sigma_{s}(\lambda) n_{s}/\cos(\alpha)dh$ where $H_{tel}$ is the altitude of the telescope. 
 
 Vertical profiles of individual showers are shaped by a random process of collisions between the high-energy particles and air molecules. They exhibit strong shower-to-shower fluctuations. However,  stacking of the vertical profiles of a large number of EAS removes the fluctuations.  
 EAS signal is detected on top of random fluctuations of the Night Sky Background. With typical reflector sizes $D_{tel}\sim 4-30$~m, currently existing IACTs are able to detect the EAS initiated by CRs with energies above the energy threshold $E_{thr}$ of several hundred GeVs hitting the ground within an area  $A_{eff}\sim 10^5$~m$^2$ around the telescope \cite{aharonian_book}. This provides a rate of detection of EAS ${\cal R}_{cr}=N_{cr}(E_{thr})A_{eff}\sim 1\left[D_{tel}/10\mbox{ m}\right]^{2(\gamma_{cr}-1)}$~kHz, where we have assumed that the energy threshold scales approximately as the telescope aperture $E_{thr}\sim D_{tel}^{-2}$.   Each particular EAS triggers the readout system of the telescope if it produces more than $N_{\gamma,min}\sim 10-10^2$ photon counts in the telescope camera. The statistics of Cherenkov signal from all the EAS accumulated each second is rather high, ${\cal R}_{cr}N_{\gamma,min}\sim 10^5\left[D_{tel}/10\mbox{ m}\right]^{2(\gamma_{cr}-1)}$~ph/s.  

Presence of an atmospheric feature (cloud or aerosol layer) of an optical depth $\tau_{cl}$ distorts the cumulative vertical profile in several ways. First, it reduces the number of Cherenkov photons reaching the telescope from behind the feature by $\exp(-\tau_{cl})$. Next, it increases the energy threshold by up to $\exp(\tau_{cl})$, depending on the altitude of the feature. This is  due to the fact that the telescope detects only EAS which produce $N_{\gamma,min}$ photon counts in the camera. Higher energy of EAS triggering the telescope is needed to compensate the loss of photons from behind the cloud. Finally, scattering of Cherenkov light in the cloud increases the signal from the altitude of the cloud and facilitates detection of the signal at large off-axis angles $\alpha$, thus imitating an  increase of the telescope's FoV. Photons which would miss the telescope in the absence of the cloud, could occasionally be scattered in the direction of the telescope and in this way contribute to the EAS image, as shown in Fig. \ref{fig:principle}. For  moderately optically thick clouds, light propagation in the cloud is dominated by single scattering events. Scattered light signal arriving in the telescopes at a distance $d_{tel}$ from the EAS footprint  is proportional to the differential cross-section of scattering $d\sigma_s/d\Omega(\theta_s)$ (or, equivalently, scattering phase function) at an angle $\theta_s=\arctan(d_{tel}/H_{cl})$. In the absence of absorption of light (good approximation in the UV band) the altitude-dependent number of scattered photons  reaching a  telescope of aperture $A_{tel}$ is
\begin{equation}
\label{eq:shower_eq1}
\frac{dN_{\gamma} }{dH}=N_{Ch}\left(n_{s}\left.\frac{d\sigma_{s}}{d\Omega}\right|_{\theta_s}\frac{A_{tel}\cos^2\theta_s}{H_{cl}^2}\right)\exp\left(-\int_{H_{tel}}^{H_{cl}} \frac{\sigma_{s} n_{s}}{\cos\theta_s}dh\right)
\end{equation}
where 
\begin{equation}
\label{eq:shower_eq2}
N_{Ch}(H)=\int_{H_{cl}}^\infty \int_\Omega \frac{dN_{Ch}}{dH} \exp\left(-\int_{H_{cl}}^h \sigma_s n_s dh'\right)d\Omega dh
\end{equation}
is the amount of Cherenkov light accumulated in the EAS from the top of the atmosphere down to the altitude $H_{cl}$. The equation (\ref{eq:shower_eq1}) closely resembles the LIDAR equation \cite{winker09}\footnote{The similarity between the LIDAR and EAS light signals in the presence of clouds was first noticed in the context of UHECR observations from space, see Ref. \cite{takahashi05}.}.

\section{Numerical simulaitons}
 
\begin{figure}
\includegraphics[width=0.5\linewidth]{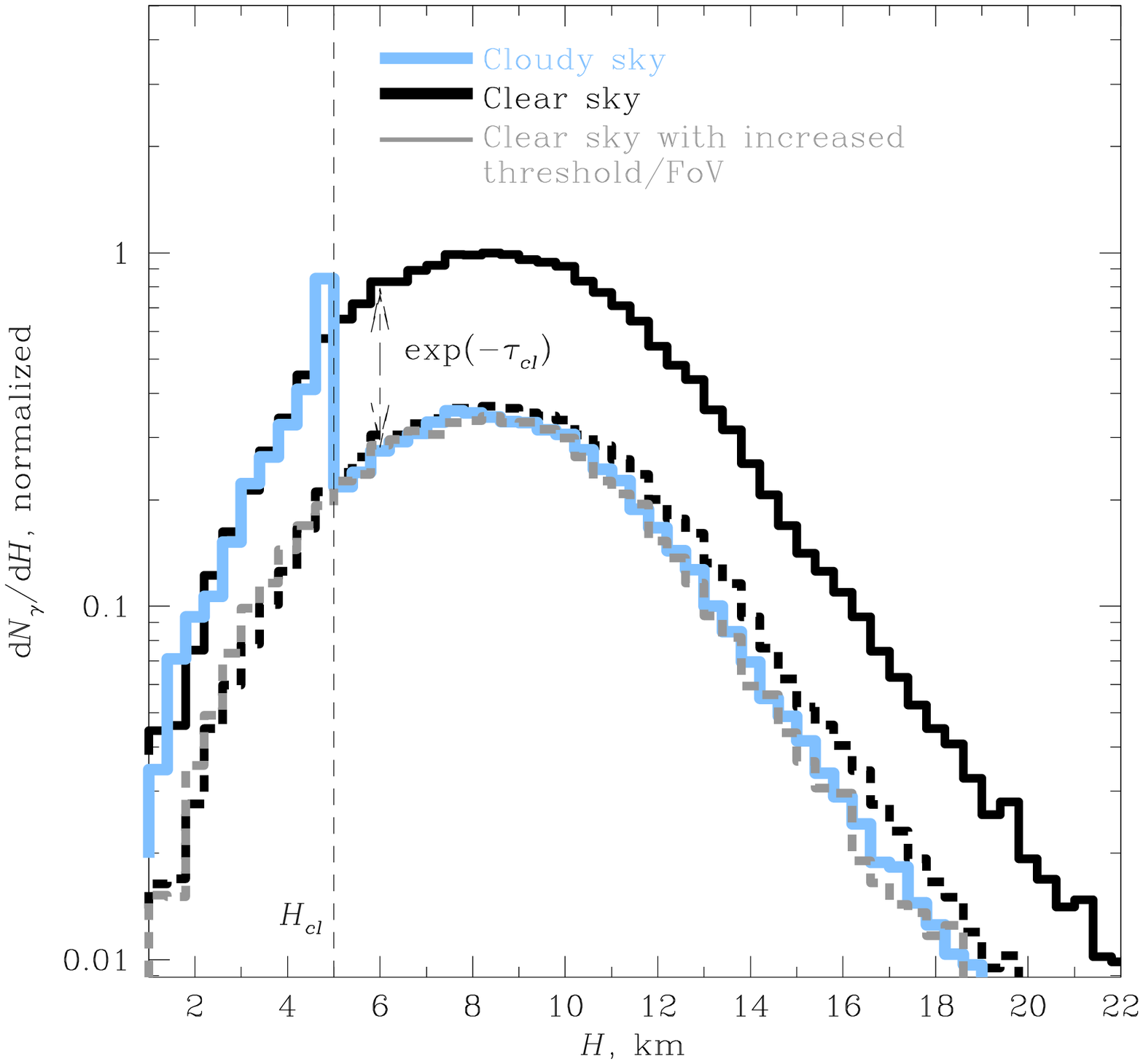}
\includegraphics[width=0.5\linewidth]{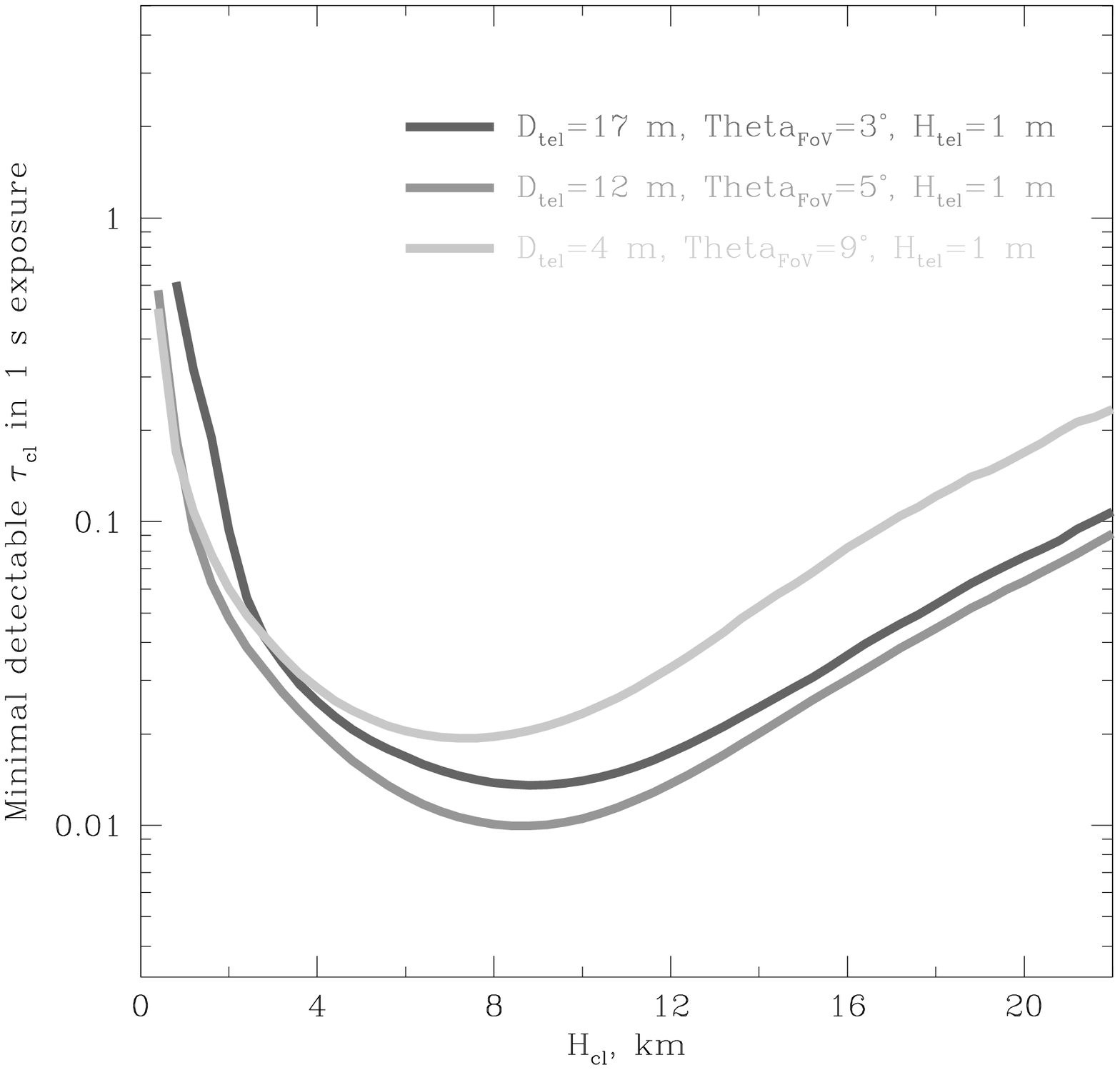}
\caption{Left: simulated vertical profiles of Cherenkov light from $t_{exp}=1$~s exposure detected by a vertically pointed IACT with dish size $D_{tel}=17$~m and FoV $\Theta_{FoV}=4^\circ$ situated at the altitude $H_{tel}=1$~m a.s.l. Light blue thick line shows the clear sky profile. Black thick line corresponds to a cloudy sky profile with a cloud at $H_{cl}=5$~km and $\tau_{cl}=1$. Thin grey lines show the clear sky profile with the adjusted threshold $N_{\gamma,min}$ and increased FoV (dashed) and normalisation decreased by $\exp(-\tau_{cl})$ (dotted). Right: minimal optical depth of an atmospheric feature detectable in 1~s exposure time by telescopes with different parameters.}
\label{fig:cloudy_sky}
\end{figure}

To verify the qualitative arguments on the influence of clouds and aerosols on the EAS signal, presented in the previous section, we have performed Monte-Carlo simulations of vertical profiles of Chernekov light in clear and cloudy sky conditions.  Our modelling  is based on Cosmic Ray Simulations for KAscade (CORSIKA) Monte-Carlo simulations of Extensive Air Showers (EAS)\footnote{http://www-ik.fzk.de/~corsika/}. We use version 7.40  of CORSIKA software. We simulate vertical EAS initiated by protons in the energy range 100~GeV -- 10~TeV following a powerlaw distribution $dN/dE\sim E^{-p}$ with the slope $p=-2.7$. 

Cherenkov emission from the EAS particles is calculated using the {\tt CHERENKOV} option of CORSIKA configuration. The simulation outputs  information on bunches of Cherenkov photons arriving at the observation level, which we arbitrarily fix to be at the altitude $H_{tel}=100$~m above sea level.  Positions and arrival directions of Cherenkov photon bunches hitting the ground within the distance up to  4~km from the EAS core are retained for the analysis. 

Telescope detecting the Cherenkov photons from the EAS is modelled as a reflector dish circle of an area $A_{tel}$ placed at a random position within 4~km from the EAS footprint.  Photons hitting the dish and arriving at an angle not larger than half-size of the telescope Field-of-View (FoV), $\Theta_{FoV}/2$ with respect to the vertical direction are counted as contributing to the EAS image. Each photon has a probability $q=0.25$, to be detected by the photosensors of the telescope camera and contribute to the EAS image. This represents typical photon detection efficiency of photomultiplier tubes or silicon photon multipliers forming the telescope camera. 

An EAS is assumed to trigger the telescope if the number of photons contributing to the image is higher than $N_{\gamma,min}=50$. The cumulative vertical profiles of Cherenkov light, shown in the left panel of Fig. 2 are calculated by adding all the Cherenkov photons from all the EAS triggering the telescope within a certain exposure time ($t_{exp}=1$~s in Fig. \ref{fig:cloudy_sky}). 

The study of the effect of the clouds and aerosol layers on the cumulative vertical profile of Cherenkov light is limited to the case of a geometrically thin cloud with moderate optical depth, $\tau_{cl}\simeq 1$. The cloud affects the Cherenkov signal in several ways. First, it suppresses the direct Cherenkov flux from the altitudes above the cloud height $H_{cl}$ by $\exp(-\tau_{cl})$. This effect is directly taken into account in the Monte-Carlo simulations. For every EAS, the number of photons in each bunch of Cherenkov photons originating from the altitude $H>H_{cl}$ is reduced by $\exp(-\tau_{cl})$ at the ground level. Second effect of the cloud on the Cherenkov signal is the enhancement of the flux from the cloud itself, due to the scattering of the photons in the cloud. We take this effect into account in the following way. For each bunch of $N_b$ photons originating from above the cloud, we generate a new bunch originating from the altitude $H_{cl}$ and containing $(1-\exp(-\tau_{cl}))N_b$ photons. The direction of the new bunch differs from the direction of the original bunch by the scattering angle $\theta_s$, which is a random number drawn from a distribution determined by the scattering phase function. In our model example we consider the scattering phase function of water droplets, of the type found in the Mie scattering theory \cite{bohren} for spheres of the size some 10~times larger than the photon wavelength: 
\begin{equation}
P(\theta_s)\simeq A_0\exp(-\theta_s/\Theta_0)+A_1\exp(-\theta_s/\Theta_1)+A_2\exp\left(-\left[\frac{(\theta_s-\Theta_2)}{\Delta\Theta_2}\right]^2\right)+A_3.
\end{equation}
Here the forward peak is determined by parameters $A_0=10^3, \Theta_0=0.3^\circ$, the broader forward wing is described by 
$A_1=15, \Theta_1=13^\circ$, and backward scattering is described by $A_2=0.13, A_3=10^{-2}, \Theta_2=145^\circ$. 

The strength of the enhancement of the signal from the cloud altitude strongly depends on the details of the shape of the scattering phase function $P(\theta_s)$, and on the distance of the shower footprint from the telescope position. 

In our model calculations we make a simplifying assumption that photon propagation in the cloud is dominated by single scattering events. This is, in general, not true for very optically thick clouds $\tau_{cl}>1$, in which multiple scattering becomes more and more important with the increase of the optical depth. Multiple scattering effects also affect the strength of the "cloud peak" feature visible in Fig. \ref{fig:cloudy_sky}. 

 Fig. \ref{fig:cloudy_sky} shows a numerically calculated stacked vertical profile of Cherenkov light from the a set of nearly vertical EAS detected in one second exposure  by a large reflector telescope (dish size $D_{tel}=17$~m) with a FoV $\Theta_{FoV}=4^\circ$ in clear and cloudy sky conditions.  The three types of distortion of the profile are clearly seen in example shown in Fig. \ref{fig:cloudy_sky}, which shows a comparison of the profile in the presence of a cloud with optical depth $\tau_{cl}=1$ at an altitude $H_{cl}=5$~km with the clear sky vertical profile of Cherenkov light.

\section{Retrieval of information on clouds and aerosols  from the Cherenkov light profile}
  
 Taking into account uncertainties related to the strength of the "cloud peak" feature in the vertical profile of Cherenkov light, our algorithm of the measurement of the cloud optical depth does not rely on the measurement of the properties of the feature. Instead, it is based on comparison of the Scattering Ratios $SR$ below and above the cloud.  

We illustrate the method of measurement of cloud /  aerosol characteristics using the example profile shown in Fig. \ref{fig:cloudy_sky}.  The shape (but not normalisation) of cloudy sky vertical profile $\left.dN_\gamma/dH\right|_{cloudy}$ follows the clear sky vertical profile $\left.dN_\gamma/dH\right|_{clear}$ in the altitude ranges $H<H_{cl}$ and $H>H_{cl}$ with properly adjusted  $N_{\gamma,min}$ and $\Theta_{FoV}$ (grey lines in Fig. \ref{fig:cloudy_sky}).  Introducing a "scattering ratio" $SR(H)=\left[\left.dN_\gamma/dH\right|_{cloudy}\right]/\left[\left.dN_\gamma/dH\right|_{clear}\right]$ we could identify the clear sky altitude ranges in which $SR(H)=const$ for some  $N_{\gamma,min}, \Theta_{FoV}$, which could be found simultaneously with the clear sky intervals via fitting of the observed vertical profile.  

The extent of the cloud corresponds to the altitude range at which $SR(H)\not= const$. Identification of the altitude of the deviation of $SR$ from constant provides a measurement of $H_{cl}$. Vertical resolution of the measurement is determined by the angular resolution of the telescope $\theta_{PSF}$ and the typical distance of the telescope from the EAS footprints on the ground, $d_{tel}=\sqrt{x^2+y^2}$: $\Delta H_{cl}=(H_{cl}^2+d_{tel}^2)\theta_{PSF}/d_{tel}$.  

Optical depth of the feature could be measured from the $SR$ below and above the feature: $\exp(-\tau_{cl})=\left[SR(H>H_{cl})\right] / \left[SR(H<H_{cl})\right]$, as it is shown in Fig. \ref{fig:cloudy_sky}. The minimal optical depth measurable in this way is determined by the statistics of the Cherenkov signal. The distortion of the profile should be larger than the statistical uncertainty estimated from the overall number of photons in the profile, $N_{\gamma,cum}={\cal R}_{cr}N_{\gamma,min}t_{exp}$, where $t_{exp}$ is the exposure time.  For a low optical depth cloud, the decrease of the signal is about $\Delta N_\gamma\sim \tau_{cl}N_\gamma$. Comparing this to the level of fluctuations of the signal, $N_\gamma^{1/2}$, one could find that the minimal detectable optical depth is $\tau_{cl,min}\sim N_\gamma^{-1/2}\sim 10^{-2}\left[D_{tel}/10\mbox{ m}\right]^{1-\gamma_{cr}}\left[t_{exp}/1\mbox{ s}\right]^{-1/2}$. 

The right panel of Fig. \ref{fig:cloudy_sky} shows the minimal detectable optical depth of a cloud as a function of the cloud altitude. It is determined by the condition that this distortion of the vertical profile of Cherenkov light above and below the cloud should be stronger than random fluctuations of the clear sky profile. Measurement of the optical depth is done via comparison of the overall normalisations of the vertical profiles above and below the cloud. The accuracy of the measurement of both normalisations is determined by the statistics of the signal from above and below the cloud, $N_{H>H_{cl}}$,  $N_{H<H_{cl}}$. The error of the measurement of both normalisations is $\Delta N_{H>H_{cl}}/N_{H>H_{cl}}=1/\sqrt{N_{H>H_{cl}}}$,   $\Delta N_{H<H_{cl}}/N_{H<H_{cl}}=1/\sqrt{N_{H<H_{cl}}}$.  The error of the measurement of the ratio of normalisations is  $\Delta\left(\exp(-\tau_{cl})\right)\simeq \Delta\tau_{cl}=\sqrt{(\Delta N_{H>H_{cl}}/N_{H>H_{cl}})^2+(\Delta N_{H<H_{cl}}/N_{H<H_{cl}})^2}$.  The right panel of Fig. 2 shows the minimal detectable optical depth of the cloud, derived from the condition $\tau_{cl}=5\Delta\tau_{cl}$. The minimal detectable optical depth is shown as a function of the altitude of the cloud for different telescope configurations. We have taken the telescopes with parameters close to those of existing and planned IACT facilities, such as HESS (telescopes with 12~m dishes and $5^\circ$ FoV)\footnote{http://www.mpi-hd.mpg.de/hfm/HESS}, MAGIC (17~m telescopes with $3^\circ$ FoV)\footnote{https://magic.mpp.mpg.de} and the telescopes of the Small-Size Telescopes (SST) sub-array of the planned CTA facility (4~m dishes, $9^\circ$ FoV) \footnote{http://www.cta-observatory.org}.  

{Longer exposure  time should allow detection of the clouds with lower optical depth and also characterisation of clouds with a more complicated vertical structure (compared to a single layer geometrically thin cloud considered in our model example). The improvement of the minimal measurable $\tau_{cl}$ with exposure time would, however, stop as soon as the level of variations of the vertical profile induced by the presence of the cloud will become comparable to the level of the systematic uncertainties of the knowledge of the instrument characteristics and their time variability.}

The scattered light signal is responsible for the enhancement of emission from the altitude of the cloud, visible in the left panel of Fig. \ref{fig:cloudy_sky}. Measurement of the strength of the scattered light peak as a function of the distance $d_{tel}$ provides a probe  of the scattering amplitude as a function of $\theta_s$ or, in other words, of the scattering phase function of the cloud particles.

 The spectrum of Cherenkov emission $dN_{Ch}/d\lambda\sim \lambda^{-2}\left(1-c^2/([n(\lambda)]^2v^2)\right)$ ($n$ is the refraction index of the air, $v$ is particle velocity) \cite{longair} is a continuum stretching through UV and visible bands. Cherenkov light provides a "white light"  source in the atmosphere. The use of such white light  for remote sensing has certain advantages compared to typically mono-wavelength light used in LIDARs \cite{kasparian03}. Namely, the broad range of the Cherenkov light opens a possibility for a measurement of the  optical depth  as a function of wavelength. This provides a tool for the measurement of sizes of aerosol particles $r_p$, since the scattering / extinction cross-section $\sigma_s(\lambda)$ depends on the size parameter $x=2\pi r_p/\lambda$. Modifications of existing IACTs are needed for such a study, because currently existing systems measure only the intensity of Cherenkov light integrated over a single spectral window of light sensors. 

\section{Conclusions}
 
 In this paper we have proposed a novel approach for the remote sounding of the atmosphere using the UV Cherenkov light generated by the cosmic ray induced EAS throughout the atmospheric volume. This approach allows detection of atmospheric features, such as cloud and aerosol layers, and characterisation of their geometrical and optical properties. 
 
{Noticing  an analogy between  the UV light pulse produced by the EAS and the pulse of the laser light, commonly used in the LIDAR devices, we demonstrated that the principles of the measurement of the properties of clouds and aerosols based on the imaging and timing of the EAS signal are very similar to those used by the LIDAR. In fact, the equations (\ref{eq:shower_eq1}), (\ref{eq:shower_eq2}) are the direct analogs of the well-known "LIDAR equation" commonly used in the analysis and interpretation of the LIDAR data. There are, however, important differences between the EAS and laser light pulses, which make the new approach based on the EAS light complementary to the LIDAR approach. Most importantly, the Cherenkov light is continuously "regenerated" all along the EAS track from the top to the bottom of the atmosphere, while the laser light is generated once in a single location (e.g. at  the ground level for the ground based LIDAR). Another important difference is that the Cherenkov light has a continuum spectrum spanning through the visible and UV bands, while the laser light of the LIDARs is mono-wavelength. We have shown that the difference in the properties of the light used by the LIDARs and by the proposed EAS + Cherenkov telescope setup potentially provides new possibilities for the measurement of physical characteristics of the cloud / aerosol particles, such as e.g. size distribution  and the scattering phase function. Thus, the proposed technique is expected to  provide data useful in the context of atmospheric physics.}
 
 {Existing IACT systems use a range of atmospheric monitoring tools to characterise weather conditions at their observation sites, including infrared / visible cameras and conventional LIDARs. The Atmospheric monitoring data are collected with the aim to control the quality of the astronomical gamma-ray data, which are the data on the UV Cherenkov emission from the EAS induced by gamma-rays coming from high-energy astronomical sources. We have demonstrated that the IACTs themselves could serve as powerful atmospheric monitoring tools, providing the atmospheric data complementary to those of the LIDARs and visible / infrared cameras. The atmospheric sounding data could be partially extracted from the background cosmic ray data of \gr\ observations by existing IACTs. Their collection does not require interruptions of  the planned astronomical observation schedule.  Moreover, the atmospheric data could be collected in cloudy sky conditions when astronomical observations are difficult or impossible.  Availability of detailed simultaneous atmospheric sounding data should allow a better control of the quality of the astronomical \gr\ data taken by existing IACTs, e.g. via a better definition of the "clear sky" conditions. Besides, this should also open a possibility for observations  in a borderline situation of the presence of moderately optically thin clouds and aerosols. }

\end{document}